\documentclass[prd,aps,eqsecnum,showpacs]{revtex4}

\usepackage{amssymb}
\usepackage{bm}

\def\prez{{}_{(0)}}

\def\pret{{}_{(2)}}
\def\preth{{}_{(3)}}

\begin{document}

\title{Gradient expansion approach to nonlinear superhorizon perturbations}

\author{Yoshiharu {\scshape TANAKA} and Misao {\scshape SASAKI}
}

\affiliation{
Yukawa Institute for Theoretical Physics, Kyoto University, 
Kyoto 606-8502, 
Japan
}


\begin{abstract}
Using the gradient expansion approach, we formulate a nonlinear 
cosmological perturbation theory on super-horizon scales 
valid to $O(\epsilon^2)$, where $\epsilon$ is the expansion parameter
associated with a spatial derivative.
For simplicity, we focus on the case of a single perfect fluid, 
but we take into account not only scalar but also vector and
tensor modes.
We derive the general solution under the uniform-Hubble 
time-slicing. In doing so, we identify the scalar, vector and
tensor degrees of freedom contained in the solution.
We then consider the coordinate transformation
to the synchronous gauge in order to compare our result with the previous
result given in the literature. In particular, we find that
the tensor mode is invariant to $O(\epsilon^2)$ under the 
coordinate transformation.
\end{abstract}

\hfill YITP-06-69

\maketitle

\section{Introduction}
Recently, there have been a number of studies investigating
the possibility of
non-Gaussian primordial curvature perturbations from inflation.
The cosmic microwave background (CMB) temperature anisotropy 
was found by COBE to be of order $10^{-5}$~\cite{COBEDMR},
and a recent more accurate observation by WMAP was found 
to be consistent with the predictions of
the inflationary universe, namely that the universe is spatially flat
and that the primordial curvature perturbation is almost
scale-invariant and statistically Gaussian~\cite{WMAP3y,WMAPng}.
Thus, it seems that the use of the linear cosmological perturbation theory
that has been developed in the last couple of
 decades~\cite{Bardeen:1980kt,Kodama:1985bj,Mukhanov:1990me}
has been observationally justified.

Nevertheless, the constraint obtained by WMAP on the deviation
from a Gaussian distribution is still very weak,~\cite{WMAPng},
and the next generation of observations, i.e. those to be carried by the PLANCK satellite, may be able
to detect a deviation from a Gaussian distribution and constrain the level of 
such deviation
 by an order of magnitude better than the WMAP observation, with only 
temperature anisotropies~\cite{Komatsu:2001rj}, and with a combination of 
temperature and polarization (only E mode) anisotropies~\cite{Babich:2004yc}.
 Furthermore, it has recently been suggested that by observing 21 cm background anisotropies, it might be possible to constrain the deviation from a 
Gaussian distribution of the primodial curvature perturbations
to an accuracy which is better than the WMAP observation by three orders of
magnitude~\cite{Cooray:2006km}. Also, theoretically,
although the deviation from Gaussian behavior obtained
from the standard single-field slow-roll
inflation is too small to be detected~\cite{Acquaviva:2002ud,Maldacena:2002vr},
many new types of inflationary models have been proposed in the last 
few years, motivated mainly by string theory and higher dimensional gravity
theories, that predict rather large deviations~\cite{NGmodels1,NGmodels2}.
Consequently, there have recently been many studies of the deviation from Gaussian behavior in inflation~\cite{Bartolo:2004if}.

The reason that a large deviation from a Gaussian distribution has not
yet been ruled out by observations lies in
the smallness of the primordial 
curvature perturbation amplitude. This is demonstrated explicitly
using the so-called $f_{NL}$ parameter introduced by
 Komatsu and Spergel~\cite{Komatsu:2001rj},
\begin{eqnarray}
\Phi(\bm{x})=\Phi_L(\bm{x})
+f_{NL}\left(\Phi_L^2(\bm{x})-\langle\Phi_L^2\rangle\right),
\label{fnl}
\end{eqnarray}
where $\Phi$ is the curvature perturbation on the Newton (or longitudinal)
slicing and $\Phi_L$ is the linear Gaussian part of the perturbation.
Because $\Phi_L\sim 10^{-5}$, the coefficient $f_{NL}$ can be
very large, even if $\Phi$ is very small.

Thus, to quantify the deviation from a Gaussian distribution and 
clarify its observational
effect, it is important to develop a theory that can treat
nonlinear cosmological perturbations. There are two approaches
to nonlinear perturbations. One is the second-order perturbation
theory~\cite{Matarrese:1997ay,SecondOrder}, which has been used by
several authors to quantify the 
non-Gaussianity~\cite{Acquaviva:2002ud,Maldacena:2002vr,NGmodels1,Seery:2005wm}.
The other employs a gradient expansion by assuming that the spatial
derivative is sufficiently small compared to the time
 derivative~\cite{Lyth:2004gb}. 
In this paper, we use the gradient expansion approach and
formulate cosmological perturbations to
full nonlinear order in its amplitude, but to second order
in spatial gradients. 

In passing, we note that while we work with the standard metric perturbation,
there exists a different
approach, called the covariant approach~\cite{Ellis:1989jt}. 
In particular, recently Langlois and Vernizzi have succeeded in
formulating it in a form that may be useful when dealing with
nonlinear perturbations not only on superhorizon scales but also 
on subhorizon scales~\cite{Langlois:2005ii}. 

In the (spatial) gradient expansion approach, we introduce an expansion
parameter, $\epsilon$, and associate it with each spatial
derivative, expand the field equations in terms of $\epsilon$,
and solve them iteratively order by order.
Originally, the gradient expansion was used
to explore the general behavior of the spacetime near
the cosmological singularity~\cite{Lifshitz:1963ps,Belinsky:1982pk}.
Tomita called it the anti-Newtonian approximation~\cite{antiNewton}
and used it to investigate cosmological perturbations on
superhorizon scales~\cite{Tomita:1975kj}. 
The terminology, `gradient expansion' is perhaps due to
Salopek and Bond~\cite{Salopek:1990jq}. 
They adapted it to the Hamilton-Jacobi formalism and used it to
investigate the nonlinear perturbations 
on superhorizon scales in slow-roll inflation.
Later, Comer et al.\cite{Comer:1994np}
 formulated the gradient expansion and discussed the
form of the solution to $O(\epsilon^4)$ beyond the leading
order, and Deruelle and Langlois
carried out a thorough study of nonlinear perturbations
near the cosmological singularity~\cite{Deruelle:1994iz}.

There exists an extensive literature on the gradient expansion method.
To summarize briefly, including the works mentioned above,
gradient expansions to higher orders by
iteratively solving the Einstein equations was formulated and studied
in Ref.~\cite{Lifshitz:1963ps, Belinsky:1982pk, antiNewton, Tomita:1975kj,
Muller:1989rp,Comer:1994np, Deruelle:1994iz, Iguchi, Khalatnikov:2003ac}.
However, in most of these works, the authors either focused on
the asymptotic behavior in the past or in the future or assumed a
specific form of the solution.
This obscures the physical degrees of freedom associated with the solutions
they obtained. In particular, no clear analysis concerning
the possible decomposition of a solution into the
scalar, vector and tensor modes has been carried out.

The Hamilton-Jacobi approach,
in which one assumes the form of the generating functional to find solutions,
was adopted
in Ref.~\cite{Salopek:1990mp, HJHigherOrder, Soda:1995fz, Nambu:1994hu}.
However, because of the simple form of the generating functional that the authors of these works assumed, the solutions they obtained were not general enough.
Specifically, they ignored half of the physical degrees of freedom
that correspond to decaying modes in the linear theory.

Perturbations on superhorizon scales have been
 studied extensively with the so-called separate
universe approach~\cite{Sasaki:1998ug,Wands:2000dp}.
The fact that the separate universe approach is essentially
the leading-order approximation of the gradient expansion was
demonstrated by Rigopoulos and Shellard~\cite{Rigopoulos:2003ak},
and a more general analysis was given by 
Lyth, Malik and Sasaki~\cite{Lyth:2004gb}.
Although this leading-order approximation seems to be
valid for quite a large number of inflation models, 
it can miss important points in some cases.
A good such example is the case studied by Leach et al.~\cite{Leach:2001zf}.
They considered the linear perturbation of a single-field
inflation model, which has a stage at which the slow-roll conditions are
violated. They then showed that the $O(k^2)$ corrections to the
curvature perturbation on superhorizon scales, where $k$ is the comoving
wavenumber of the perturbation, play a crucial role in
the determination of the final curvature perturbation amplitude.
More precisely, it is the decaying mode solution which
becomes non-negligible and makes the $O(k^2)$ corrections important.
If the $O(k^2)$ corrections are important already at linear order,
one expects that they will significantly affect the deviation from 
a Gaussian behavior.
In the gradient expansion, these corrections correspond to $O(\epsilon^2)$
beyond the leading order.
Thus, it is important to include the $O(\epsilon^2)$ terms
as well as the decaying modes in the gradient expansion.

Here, we mention an another reason for the necessity of going beyond the 
leading-order approximation.
While nothing remains unclear about the gauge dependence 
and/or gauge invariance in the linear perturbation theory,
there are still some issues to be clarified in the case of
nonlinear, superhorizon perturbations. Among them are the issue of
identifying the scalar, vector and tensor modes and their gauge 
(or coordinate) transformation properties. Specifically,
tensor modes which are to describe gravitational
waves are not clearly seen at leading order. This is
because of the general covariance according to which,
even if there exits a gravitational wave, spacetime can be 
made locally Minkowski within a region much smaller 
than the wavelength of the gravitational wave.
Thus, it is necessary to go to higher orders to
clearly distinguish tensor, scalar and vector modes
and to clarify the gauge dependence/invariance.

In connection with this, we note that most authors have
worked in the synchronous gauge.
Of course nothing is particularly wrong with
the synchronous gauge, but it is not the most convenient choice
for the analysis of superhorizon scale perturbations.
One apparent drawback is the existence of a gauge mode
in the time coordinate.
In other words, the synchronous time-slicing
condition does not fix the slices uniquely.
Thus, it is worth developing the gradient expansion approach
in an appropriate time-slicing which fixes the time coordinate
 uniquely, in order to understand the physical meaning of solutions.

In summary, there are several important issues in the
gradient expansion approach which need more
detailed investigations: derivation of 
the general solution in the sense of the Cauchy initial data;
identification of the scalar, vector and tensor modes, if
possible, in the general solution; and clarification of the
gauge dependence/invariance of the general solution.

In this paper, to address these issues, 
we formulate the gradient expansion to $O(\epsilon^2)$
beyond leading order in the uniform Hubble slicing.
In the separate universe approach, or equivalently
at leading order in the gradient expansion, 
the uniform density slicing (or equivalently uniform Hubble slicing),
which fixes the time coordinate uniquely, 
is convenient~\cite{Sasaki:1998ug,Shibata:1999zs},
and a nonlinear version of a gauge-invariant combination
that corresponds to the curvature perturbation on
the uniform density slicing has been constructed~\cite{Lyth:2004gb}.
This quantity is independent of the choice of time-slicing and
conserved for adiabatic perturbations. Thus it
represents a nonlinear scalar mode perturbation.
This fact motivates us to adopt the uniform Hubble slicing.

We employ the $(3+1)$-decomposition
(Hamiltonian formalism) of the Einstein equations.
For simplicity, we assume a perfect fluid 
equation of state, $P=(\Gamma-1)\rho$, with $\Gamma$ a constant.
We then derive the general solution for all the variables,
focusing particularly
on the correspondence between the degrees of freedom appearing
in the general solution and those in the linear theory.
We find that the identification of the tensor mode in the
spatial metric is rather arbitrary, depending on the choice of 
the spatial coordinates, as a reflection of the general
covariance, while it can be unambiguously identified in
the extrinsic curvature of the metric, albeit non-locally.

Next, we consider a coordinate transformation from the
uniform Hubble slicing to the synchronous slicing in order to compare 
our result with the results obtained in the synchronous gauge.
The main purpose of this is to clarify the relation between the variables
defined in each coordinate system. But this is also useful as 
a good way to confirm the computation. 
For definiteness, we compare our result with that
obtained by Comer et al.~\cite{Comer:1994np}.
We find that they are identical, provided that we choose
the form of the metric in accordance with their assumption.
Furthermore, we find the tensor mode is invariant
under the change of the time-slicing. In this connection, we note that 
Matarrese et al. discussed the effect of a gauge transformation
in the second-order perturbation theory in the Einstein-de Sitter
universe~\cite{Matarrese:1997ay}. Comparison of their result
with ours is an interesting problem left for future study.

This paper is organized as follows. In Section II, we define the basic
variables we use in our paper using the $(3+1)$ decomposition.
In Section III, we first state the basic assumptions in the gradient expansion.
Then we derive the general solution on the uniform Hubble slicing
to $O(\epsilon^2)$ beyond leading order, and identify the 
scalar, vector and tensor modes in the general solution.
In Section IV, we consider the transformation from the
uniform Hubble slicing to the synchronous slicing, and
clarify the relation between the result obtained by Comer et al.~\cite{Comer:1994np}
and our solution.
We conclude the paper in Section V.
Some useful order-counting formulas are given in the appendix.
\section{Basic equations}
\label{sec:basiceq}

In the $(3+1)$-decomposition, the metric is expressed as
\begin{eqnarray}
ds^2&=&g_{\mu\nu}dx^{\mu}dx^{\nu}
\nonumber \\
&=&(-\alpha^2+\beta_k\beta^k)dt^2+2\beta_idx^idt+\gamma_{ij}dx^idx^j,
\label{eq:metric}
\end{eqnarray}
where $\alpha$, $\beta^i$ ($\beta^i=\gamma^{ij}\beta_j)$, 
and $\gamma_{ij}$ are the lapse function, shift vector, and 
the 3-dimensional spatial metric, respectively.
 We rewrite $\gamma_{ij}$ as
\begin{eqnarray}
\gamma_{ij}(t,x^k)=a^2(t)\,\psi^4(t,x^k)\,\tilde{\gamma}_{ij}(t,x^k)\,;
\quad \det(\tilde{\gamma}_{ij})=1\,,
\label{eq:3metric}
\end{eqnarray}
where the function $a(t)$ is the scale factor of a fiducial
homogeneous and isotropic background universe.

The extrinsic curvature $K_{ij}$ of a hypersurface of constant $t$
is defined by
\begin{eqnarray}
K_{ij}\equiv -\nabla_in_{j}=\alpha\Gamma^0{}_{ij}\,,
\label{Kdef}
\end{eqnarray}
where $n_\mu=(-\alpha,0,0,0)$ is the vector unit normal 
to the time slices. We decompose the extrinsic curvature as
\begin{equation}
K_{ij}=\frac{\gamma_{ij}}{3}K+\psi^4a^2\tilde{A}_{ij}\,;
\quad K\equiv \gamma^{ij}K_{ij}\,,
\label{eq:Kij}
\end{equation}
where $\tilde{A}_{ij}$ represents the traceless part of $K_{ij}$.
The factors in front of $\tilde{A}_{ij}$ are determined so that
the mixed components $K^i{}_j$ take the form
\begin{eqnarray}
K^i{}_j=\frac{1}{3}\delta^i_j K+\tilde A^i{}_j\,,
\end{eqnarray}
where the indices of $\tilde{A}_{ij}$ are raised and
lowered by $\tilde{\gamma}^{ij}$ and $\tilde{\gamma}_{ij}$.

 The stress-energy tensor for the fluid is 
\begin{equation}
T_{\mu\nu}=(\rho+P)u_{\mu}u_{\nu}+Pg_{\mu\nu},
\label{eq:emtensor}
\end{equation}
where $\rho$, $P$ and $u^{\mu}$
are the energy density, pressure, and the 4-velocity, respectively.
We consider a perfect fluid with the barotropic
 equation of state $P/\rho=\Gamma-1=$const.
 The components of the 4-velocity are given by
\begin{eqnarray}
&&u^0=[\alpha^2-(\beta_k+v_k)(\beta^k+v^k)]^{-1/2},
\quad u^i=u^0v^i ,\quad v^i\equiv\gamma^{ij}v_j,
\nonumber \\
&&u_0=-u^0[\alpha^2-\beta^k(\beta_k+v_k)],
\quad u_i=u^0(v_i+\beta_i).
\label{eq:4velocity}
\end{eqnarray}
We express the $(3+1)$-decomposition of the energy-momentum tensor as
\begin{eqnarray}
&&E\equiv T_{\mu \nu}n^{\mu}n^{\nu}=(\rho+P)(\alpha u^0)^2-P,
\label{eq:Edef} 
\\
&&J_j\equiv -T_{\mu \nu}n^{\mu}\gamma^{\nu}_j=(\rho+P)\alpha u^0 u_j,
\label{eq:Jdef} \\
&&S_{ij}\equiv T_{ij}=(\rho+P)(u^0)^2(v_i+\beta_i)(v_j+\beta_j)
+P\gamma_{ij}, 
\label{eq:Sdef} \\ 
&&S^k{}_k\equiv \gamma^{kl}S_{lk} .
\label{eq:TrSdef}
\end{eqnarray}

The hydrodynamic equations $\nabla^{\mu}T_{\mu\nu}=0$ 
are written in the form~\cite{Shibata:1999zs}
\begin{eqnarray}
&&\partial_t(w\psi^6 a^3 \rho^{1/\Gamma})
+\partial_k(w\psi^6a^3 \rho^{1/\Gamma}v^k)=0,
\label{eq:5} \\  
&&\partial_t\{w\psi^6a^3(\rho+P)u_j\}+\partial_k\{w\psi^6a^3(\rho+P)
v^k u_j\}
\nonumber \\
&& \quad =-\alpha \psi^6 a^3 \partial_j P+w\psi^6a^3(\rho+P)
\{-\alpha u^0\partial_j \alpha +u_k \partial_j \beta^k
-\frac{u_k u_l}{2u^0}\partial_j \gamma^{kl}\},
\label{eq:6}
\end{eqnarray}
where $w\equiv \alpha u^0$, and we have
\begin{equation}
v^k=-\beta^k+\tilde{\gamma}^{kl}\frac{u_l}{\psi^4a^2u^0},
\label{eq:3velocity}
\end{equation}
from Eq.~(\ref{eq:4velocity}).

In the $(3+1)$-formalism of the Einstein equations, the
dynamical variables are $\gamma_{ij}$ and $K_{ij}$.
The $(n,n)$ and $(n,j)$ components of the Einstein equations
give the Hamiltonian and momentum constraint equations, respectively,
while the $(i,j)$ components gives the evolution equations for $K_{ij}$. 
The evolution equations for $\gamma_{ij}$ are given by
the definitions of the extrinsic curvature~(\ref{Kdef}).

In the present case, the Hamiltonian and momentum constraints
are
\begin{eqnarray}
&&R-\tilde{A}_{ij}\tilde{A}^{ij}+\frac{2}{3}K^2=16\pi GE\,,
\label{eq:hamconst} 
\\
&& D_i \tilde{A}^i{}_j-\frac{2}{3}D_jK=8\pi GJ_j \,.
\label{eq:momconst}
\end{eqnarray}
The evolution equations for $\gamma_{ij}$ are given by
\begin{eqnarray}
&&(\partial_t-\beta^k\partial_k)\psi+\frac{\dot{a}}{2a}\psi
=\frac{\psi}{6}\{-\alpha K+\partial_k\beta^k \},
\label{eq:dotpsi} \\
&&(\partial_t-\beta^k\partial_k)\tilde{\gamma}_{ij}=-2\alpha  \tilde{A}_{ij}
+\tilde{\gamma}_{ik} \partial_j\beta^k+\tilde{\gamma}_{jk}\partial_i\beta^k
-\frac{2}{3}\tilde{\gamma}_{ij}\partial_k\beta^k\,,
\label{eq:dotgamma}
\end{eqnarray}
where the dot represents $d/dt$.
The evolution equations for $K_{ij}$ are given by
\begin{eqnarray}
(\partial_t-\beta^k\partial_k)K&=&\alpha(\tilde{A}_{ij}\tilde{A}^{ij}
+\frac{1}{3}K^2)-D_kD^k\alpha+4\pi G\alpha(E+S^k{}_k),
\label{eq:dotK} \\
(\partial_t-\beta^k\partial_k)\tilde{A}_{ij}&=&\frac{1}{a^2\psi^4}
[\alpha(R_{ij} -\frac{\gamma_{ij}}{3}R)-(D_iD_j\alpha-\frac{\gamma_{ij}}{3}D_kD^k\alpha)]
\label{eq:dotA} \\ 
& &
+\alpha(K\tilde{A}_{ij}-2\tilde{A}_{ik}\tilde{A}^k{}_j)
+\tilde{A}_{ik}\partial_j\beta^k+\tilde{A}_{jk}\partial_i\beta^k
\nonumber  \\
& &
-\frac{2}{3}\tilde{A}_{ij}\partial_k\beta^k
-\frac{8\pi G\alpha}{a^2\psi^4}(S_{ij}-\frac{\gamma_{ij}}{3}S^k{}_k),
\nonumber 
\end{eqnarray}
where $R_{ij}$ is the Ricci tensor of the metric $\gamma_{ij}$, with
$R\equiv \gamma^{ij}R_{ij}$, and $D_i$ is the covariant derivative 
with respect to $\gamma_{ij}$.

We define the local Hubble parameter as $1/3$ of the expansion of
the unit normal vector $n^\mu$, which is
equal to $-1/3$ of the trace of the extrinsic curvature
in our convention:
\begin{equation}
3H\equiv -K=\frac{3\dot{a}}{\alpha a}+6\frac{\partial_t{\psi}}{\alpha\psi}
-\frac{D_i \beta^i}{\alpha} \,.
\label{eq:Hubbledef}
\end{equation}
In the following, we adopt the uniform Hubble slicing.
For this slicing, we have
\begin{equation}
H(t)=\frac{\dot{a}}{a}\,,
\label{eq:uniHubble}
\end{equation}
and Eq.~(\ref{eq:Hubbledef}) implies
\begin{equation}
\alpha
=\displaystyle1+\frac{2\partial_t{\psi}}{H\psi}-\frac{D_i\beta^i}{3H}
\,.
\label{eq:alpha}
\end{equation}

\section{Gradient expansion}
\label{sec:gradexpand}

\subsection{Assumptions}
We consider nonlinear superhorizon perturbations
in the approach called the anti-Newtonian 
approximation \cite{Tomita:1975kj}, the spatial gradient
 expansion \cite{Salopek:1990jq}, or the long wavelength
 approximation \cite{Shibata:1999zs}. In this approach,
we assume that the characteristic
 length scale $L$ of inhomogeneities is always much larger than the
 Hubble horizon scale, i.e. $L\gg H^{-1}\sim t$. We introduce a small parameter
 $\epsilon$, and assume that $L$ is $O(1/\epsilon)$. 
This assumption is equivalent to assuming that the magnitude of spatial
gradients of the quantities is given by
$\partial_i \psi=\psi\times O(\epsilon)$,
$\partial_i \alpha=\alpha\times O(\epsilon)$, and so on. In the limit
 $L\rightarrow\infty$, i.e., $\epsilon\rightarrow 0$, the universe
looks locally like a FLRW spacetime, where 'locally' means as seen
on the scale of the Hubble horizon volume. 
It is noted that physical quantities which are approximately
homogeneous on each Hubble horizon scale can vary nonlinearly
on very large scales. Thus this approach at leading order
is called the separate universe hypothesis~\cite{Sasaki:1998ug, Wands:2000dp}.

The local homogeneity and isotropy imply that anisotropic
 quantities, i.e., $\beta^i$ and $v^i$ should be at least $O(\epsilon)$.
Also, for the spatial metric to be locally homogeneous and isotropic,
 $\tilde{\gamma}_{ij}$ should be time 
independent in the limit $\epsilon\to0$~\cite{Lyth:2004gb}.
This implies $\partial_t{\tilde{\gamma}}_{ij}=O(\epsilon)$.
As discussed in \S~\ref{sec:gradexpand}.B,
allowing it to be $O(\epsilon)$ introduces a decaying mode
at this order, which affects all the quantities at $O(\epsilon^2)$
and makes the analysis too complicated. 
However, as discussed there, 
we have $\partial_t{\tilde{\gamma}}_{ij}=O(\epsilon^N)$
with $N>3$ in the case of an inflationary universe.
Therefore, we expect that requiring a bit tighter condition,
$\partial_t{\tilde{\gamma}}_{ij}=O(\epsilon^2)$,
will not make our analysis meaningless.
In other words, the assumption
$\partial_t{\tilde{\gamma}}_{ij}=O(\epsilon^2)$
should be sufficiently general for our analysis to be applicable
to a sufficiently wide class of cosmological situations.
Thus our basic assumptions are
\begin{eqnarray}
\beta^i=O(\epsilon)\,,
\quad
v^i=O(\epsilon)\,,
\quad
\partial_t{\tilde{\gamma}}_{ij}=O(\epsilon^2)\,.
\label{assumptions}
\end{eqnarray}
We note that from Eq.~(\ref{eq:dotgamma}), 
the assumptions $\beta^i=O(\epsilon)$ and
$\partial_t{\tilde{\gamma}}_{ij}=O(\epsilon^2)$ imply
$\tilde{A}_{ij}=O(\epsilon^2)$. 

Next, we introduce
\begin{equation}
\delta\equiv \rho/\rho_0-1,
\label{eq:20.1}
\end{equation}
where $\rho_0$ is a fiducial homogeneous part of $\rho$, which is
defined by 
\begin{eqnarray}
\rho_0(t)=\frac{3H^2(t)}{8\pi G}\,.
\end{eqnarray}
Here, $H(t)$ is given by Eq.~(\ref{eq:uniHubble}).
Then, using the equations presented in the previous section,
we evaluate the orders of magnitude of the basic
variables as~\cite{Shibata:1999zs}
\begin{eqnarray}
&&
\psi=O(1)\,,
\quad\chi\equiv \alpha-1=O(\epsilon^2)\,,
\quad w=\alpha u^0=1+O(\epsilon^6)\,,
\nonumber\\
&&
\partial_t\tilde{\gamma}_{ij}=O(\epsilon^2)\,,
\quad \partial_t{\psi}=O(\epsilon^2)\,,
\quad \tilde{A}_{ij}=O(\epsilon^2)\,,
\nonumber \\
&&
\delta=O(\epsilon^2)\,,
\quad u_i=O(\epsilon^3)\,,
\quad \beta^i+v^i=O(\epsilon^3)\,.
\label{eq:ordercount}
\end{eqnarray}
Except for $\partial_t\tilde\gamma_{ij}=O(\epsilon^2)$,
the orders of magnitude of the above quantities can all be derived
from the Einstein equations.
An explicit demonstration of the above evaluations is given in the appendix.

In particular, the fact that $\psi$ is constant
 in time to $O(\epsilon)$ follows from the perfect fluid assumption,
$P=P(\rho)$. This corresponds to the conservation of the adiabatic
curvature perturbation in the linear theory. Hence, we may regard $\psi$
as a nonlinear generalization of the curvature perturbation, although
the actual spatial curvature $R$ is no longer described solely in terms
of $\psi$ in the nonlinear case.

We also note that the curvature perturbation is defined in several
different ways in the literature. The correspondence between $\psi$ used
here and the conventions used in two other papers are
\begin{eqnarray}
\psi^2=\exp[\,\psi_{\rm LMS}\,]=\exp[\,\zeta_{\rm Mal}\,]\,,
\end{eqnarray}
where $\psi_{\rm LMS}$ and $\zeta_{\rm Mal}$ are the quantities 
used by Lyth et al.~\cite{Lyth:2004gb} 
and Maldacena~\cite{Maldacena:2002vr}, respectively.

\subsection{The leading-order solution}
First we derive the leading-order solution for our basic variables.
We note, however, that this does not mean that we only solve the Einstein equations at lowest order in the gradient expansion. For example, from 
Eq.~(\ref{eq:ordercount}), we have $\tilde A_{ij}=O(\epsilon^2)$.
Hence we may ignore it at lowest order in the gradient expansion,
while here we do solve for the leading-order part of $\tilde A_{ij}$.

Taking the $O(\epsilon^0)$ part of Eqs. (\ref{eq:hamconst}) 
and (\ref{eq:dotK}), 
we have
\begin{eqnarray}
&&\frac{2}{3}K^2=16\pi G\rho_0,
\label{eq:20.2} \\
&&\partial_tK=\frac{1}{3}K^2+4\pi G(3\Gamma-2)\rho_0.
\label{eq:20.3}
\end{eqnarray}
Integrating these equations, we obtain
\begin{eqnarray}
&&a(t)=a_ft^{\frac{2}{3\Gamma}},
\label{eq:20.4} \\
&&\rho_0(t)=\frac{1}{6\pi G\Gamma^2 t^2},
\label{eq:20.5}
\end{eqnarray}
where $a_f$ is a constant whose normalization is arbitrary.

Substituting the order of magnitude evaluations of the variables
shown in Eq.~(\ref{eq:ordercount}) into the hydrodynamic equations
 given in \S2, 
we find
\begin{eqnarray}
&&\frac{1}{\Gamma}\partial_t{\delta}
+\frac{6\partial_t{\psi}}{\psi}+D_k v^k=O(\epsilon^4),
\label{eq:22} \\
&&\partial_t\{\psi^6\rho_0a^3\Gamma u_j\}=-\psi^6\rho_0a^3[\Gamma\partial_j\chi
+(\Gamma-1)\partial_j\delta]+O(\epsilon^5)\,.
\label{eq:23}
\end{eqnarray}

The Hamiltonian and momentum constraint equations give
\begin{eqnarray}
&&\tilde{\gamma}^{ij}\tilde{D}_i\tilde{D}_j\psi=\frac{1}{8}
\tilde{\gamma}^{kl}\tilde{R}_{kl}\psi
-2\pi G\psi^5a^2\rho_0\delta+O(\epsilon^4),
\label{eq:aphamconst} \\
&&\tilde{D}^j(\psi^6\tilde{A}_{ij})
=8\pi G\Gamma\rho_0u_i\psi^6+O(\epsilon^5),
\label{eq:apmomconst}
\end{eqnarray}
where $\tilde{R}_{ij}$ is the Ricci tensor with respect 
to $\tilde{\gamma}_{ij}$, $\tilde{D}_i$ is the covariant derivative 
with respect to $\tilde{\gamma}_{ij}$.

The evolution equations for the spatial metric give
\begin{eqnarray}
&&\frac{6\partial_t{\psi}}{\psi}-3\frac{\dot{a}}{a}\chi=D_k \beta^k,
\label{eq:25} \\
&&(\partial_t-\beta^k\partial_k)\tilde{\gamma}_{ij}=-2\tilde{A}_{ij}
+\tilde{\gamma}_{ik}\partial_j\beta^k+\tilde{\gamma}_{jk}\partial_i\beta^k
-\frac{2}{3}\tilde{\gamma}_{ij}\partial_k\beta^k+O(\epsilon^4)\,,
\label{eq:26}
\end{eqnarray}
while the evolution equations for the extrinsic curvature give
\begin{eqnarray}
&&\partial_t \tilde{A}_{ij}+3\frac{\dot{a}}{a}\tilde{A}_{ij}
=\frac{1}{a^2\psi^4}[R_{ij}-\frac{\gamma_{ij}}{3}R]+O(\epsilon^4)\,,
\label{eq:29} \\
&&\nabla^2 \chi
=4\pi G\rho_0a^2\{(3\Gamma-2)\delta+3\Gamma\chi\}+O(\epsilon^4)\,,
\label{eq:30}
\end{eqnarray}
where
\begin{eqnarray}
\nabla^2 \equiv \frac{1}{\psi^6}\partial_k(\psi^2\tilde{\gamma}^{kl}
\partial_l)\,.
\label{eq:24}
\end{eqnarray}

We now solve the above equations to derive the leading-order 
general solution for each variable.
{}From Eq.~(\ref{eq:ordercount}), we have
\begin{eqnarray}
\psi=\prez L(x^i)+O(\epsilon^2),
\label{eq:psizero}
\end{eqnarray}
where $\prez L$ is an arbitrary function of the spatial coordinates
and is $O(\epsilon^0)$. Here and in what follows, the prefix $(n)$ 
means that the quantity is $O(\epsilon^n)$.

{}From Eq.~(\ref{eq:22}) and Eq.~(\ref{eq:25}), we obtain 
\begin{equation}
\partial_t{\delta}
 + 3\Gamma\frac{\dot{a}}{a} \chi = O(\epsilon^4). 
\label{eq:32}
\end{equation}
Since the left-hand side of Eq.~(\ref{eq:30}) is $O(\epsilon^4)$,
the fact that both $\delta$ and $\chi$ are $O(\epsilon^2)$ gives
\begin{eqnarray}
\chi = -\frac{3\Gamma-2}{3\Gamma}\,\delta  + O(\epsilon^4). 
\label{eq:33}
\end{eqnarray}
{}From Eqs.~(\ref{eq:32}) and (\ref{eq:33}), we have
\begin{eqnarray}
\partial_t{\delta}-(3\Gamma-2)\frac{\dot{a}}{a} \delta = O(\epsilon^4). 
\label{eq:34}
\end{eqnarray}
Integrating this equation yields
\begin{eqnarray}
\delta=\pret Q(x^i)t^{2-\frac{4}{3\Gamma}} +O(\epsilon^4)\,.
\label{eq:delta}
\end{eqnarray}
Then, Eq.~(\ref{eq:33}) gives
\begin{eqnarray}
\chi=-\frac{3\Gamma-2}{3\Gamma}\,\pret Q(x^i)\, t^{2-\frac{4}{3\Gamma}}
 + O(\epsilon^4)\,.
\label{eq:chisol}
\end{eqnarray}

Next, we seek the general solution for $u_l$. 
First, using Eq.~(\ref{eq:33}), Eq.~(\ref{eq:23}) becomes
\begin{equation}
\partial_t\{\rho_0a^3\Gamma u_j\}=\frac{\rho_0a^3\partial_j\delta}{3}
+O(\epsilon^5).
\label{eq:36.1}
\end{equation}
Solving this equation, we obtain
\begin{equation}
u_j=\frac{\partial_j\pret Q(x^i)}{3\Gamma+2}t^{3-\frac{4}{3\Gamma}}
+\preth C_j(x^i)t^{2-\frac{2}{\Gamma}},
\label{eq:e36.2}
\end{equation}
where $\preth C_j(x^i)$ is an arbitrary vector of the spatial coordinates.

To find the leading-order solution for $\tilde A_{ij}$,
we need to evaluate the spatial Ricci tensor $R_{ij}$.
For this purpose, we first consider the form of the
spatial metric. Noting the relation $\partial_t \tilde{\gamma}_{ij}=O(\epsilon^2)$,
we define 
\begin{equation}
H_{ij}(t,x^i)\equiv \tilde{\gamma}_{ij}-\prez f_{ij}(x^i)=O(\epsilon^2),
\label{eq:Hij}
\end{equation}
where $\prez f_{ij}$ is an arbitrary function of the spatial coordinates.
Namely, $H_{ij}(t,x^i)$ is the $O(\epsilon^2)$ part of $\tilde\gamma_{ij}$.
Then the Ricci tensor of the spatial metric $\psi^4\tilde\gamma_{ij}$
can be decomposed as
\begin{equation}
R_{ij}=\tilde{R}_{ij} +R^{\psi}_{ij}\,,
\label{eq:Ricci}
\end{equation}
where $\tilde{R}_{ij}$ is the Ricci tensor with respect to
 $\tilde{\gamma}_{ij}$, and
\begin{equation}
R^{\psi}_{ij}\equiv -\frac{2}{\psi} \tilde{D}_i \tilde{D}_j \psi 
-\frac{2}{\psi} \tilde{\gamma}_{ij} \tilde{\Delta} \psi 
+\frac{6}{\psi^2} \tilde{D}_i \psi \tilde{D}_j \psi
-\frac{2}{\psi^2} \tilde{\gamma}_{ij} \tilde{D}_k \psi \tilde{D}^k \psi\,,
\label{eq:Rpsi}
\end{equation}
where $\tilde{\Delta}$ is the Laplacian with respect to $\tilde{\gamma}_{ij}$.
Using Eqs.~(\ref{eq:Hij}) and (\ref{eq:psizero}), we have 
\begin{eqnarray}
\tilde{R}_{ij}&=&\pret \bar{R}_{ij}+O(\epsilon^4),
\nonumber \\
R^{\psi}_{ij}&=&\pret R^L_{ij}+O(\epsilon^4),
\label{eq:Rpsizero}
\end{eqnarray}
where $\pret\bar{R}_{ij}$ is the Ricci tensor with respect to $\prez f_{ij}$,
and
\begin{eqnarray}
\pret R^L_{ij}&\equiv& -\frac{2}{\prez L}\bar{D}_i\bar{D}_j\prez L
-\frac{2}{\prez L}\prez f_{ij}\bar{\Delta}
\prez L
\label{eq:50}\\
&&\qquad\qquad+\frac{6}{\prez L^2}\bar{D}_i\prez L\bar{D}_j\prez L
-\frac{2}{\prez L^2}\prez f_{ij}\bar{D}_k\prez L\bar{D}^k\prez L,
\nonumber
\end{eqnarray}
where $\bar{D}_i$ is the covariant derivative with respect to $\prez f_{ij}$,
and $\bar{\Delta}$ is the Laplacian with respect to $\prez f_{ij}$.

Now, we seek the general solution for $\tilde{A}_{ij}$.
{}Using Eqs.~(\ref{eq:Ricci}) and (\ref{eq:Rpsizero}),
 the evolution equation for $\tilde{A}_{ij}$, Eq.~(\ref{eq:29}), becomes 
\begin{equation}
\partial_t \tilde{A}_{ij}+3\frac{\dot{a}}{a}\tilde{A}_{ij}
=\frac{\pret F_{ij}(x^k)}{a^2}+O(\epsilon^4),
\label{eq:Aijeq}
\end{equation}
where 
\begin{eqnarray}
&&\pret F_{ij}\equiv \frac{1}{\prez L^4}\left[\pret \bar{R}_{ij}
+\pret R^L_{ij}-\frac{\prez f_{ij}}{3}\,
\prez f^{kl}\left(\pret \bar{R}_{kl}+\pret R^L_{kl}\right)\right]\,.
\label{eq:Fijdef}
\end{eqnarray}
Equation~(\ref{eq:Aijeq}) can be immediately integrated, and we obtain
\begin{equation}
\tilde{A}_{ij}=\frac{\pret F_{ij}}{a^3}\int^t
 a(t^{\prime})dt^{\prime}+\frac{\pret C_{ij}(x^k)}{a^3}+O(\epsilon^4),
\label{eq:53}
\end{equation}
where $\pret C_{ij}$ is an arbitrary symmetric, traceless tensor
of the spatial coordinates.
Thus we obtain
\begin{equation}
\tilde{A}_{ij}=\frac{3\Gamma}{a_f^2(3\Gamma+2)}\,
\pret F_{ij}(x^k)\,t^{1-\frac{4}{3\Gamma}}
+\frac{\pret C_{ij}(x^k)}{a_f^3}t^{-\frac{2}{\Gamma}}.
\label{eq:54}
\end{equation}
 Here we note that, because the last term
proportional to $\pret C_{ij}\,a^{-3}$ is a homogeneous solution to
Eq.~(\ref{eq:Aijeq}), it may be $O(\epsilon)$
in principle. However, allowing it to be $O(\epsilon)$ alters
the $O(\epsilon^2)$ part of the Hamiltonian
constraint~(\ref{eq:aphamconst}) substantially,
because the quadratic term $\tilde A_{ij}\tilde A^{ij}$
 would no longer be negligible.
This situation is essentially the same as that in the case of homogeneous,
anisotropic cosmologies~\cite{Belinsky:1982pk,Tomita:1975kj,Deruelle:1994iz},
in which the square of the shear (the traceless part of the extrinsic curvature)acts as an effective energy density proportional to $a^{-6}$.
To avoid this complication, for the sake of simplicity,
we have assumed $\partial_t\tilde\gamma_{ij}=O(\epsilon^2)$,
which implies $\tilde A_{ij}=O(\epsilon^2)$.
We have not investigated how restrictive this additional
assumption is. However, in the case of an accelerating universe,
$0<\Gamma<2/3$, which would mimic an inflationary universe,
this decaying mode ($\propto a^{-3}$) is expected 
to be of order $H(k/(aH))^N$, where $N=3(2-\Gamma)/(2-3\Gamma)$,
so that $H(k/(aH))^N\propto a^{-3}$. Since $N>3$ for
$0<\Gamma<2/3$, it should be higher than third order in $\epsilon$. 
Thus, the assumption
$\tilde A_{ij}=O(\epsilon^2)$ should be valid.


Up to now we have not considered the Hamiltonian and momentum
constraint equations. The constraint equations will relate the
arbitrary spatial quantities $\prez L$, $\prez f_{ij}$, $\pret Q$,
$\pret C_{ij}$ and $\preth C_j$.

The $O(\epsilon^2)$ part of Hamiltonian constraint~(\ref{eq:aphamconst}) yields
\begin{eqnarray}
\pret Q&=&\frac{3\Gamma^2}{a_f^2\prez L^5}
\left[-\prez f^{ij}\bar{D}_i\bar{D}_j\prez L
+\frac{1}{8}\prez f^{kl}\pret\bar{R}_{kl}\prez L\right]+O(\epsilon^4),
\nonumber \\
&=&\frac{3\Gamma^2}{8a_f^2\prez L^4}\,
\prez f^{kl}\left[\pret R^L_{kl}+\pret \bar{R}_{kl}\right]+O(\epsilon^4)
=\frac{3\Gamma^2}{8a_f^2}\,R\left[L^4f\right]+O(\epsilon^4)\,,
\label{eq:Qdef} 
\end{eqnarray}
where $R\left[L^4f\right]$
 is the Ricci scalar of the metric $\prez L^4\prez f_{ij}$.
Thus $\pret Q$ is not arbitrary, but is expressed in 
terms of $\prez f_{ij}$ and $\prez L$.

The $O(\epsilon^3)$ part of the momentum constraint~(\ref{eq:apmomconst}) yields
\begin{eqnarray}
\bar{D}^j\left[\prez L^6\tilde{A}_{ij}\right]
=8\pi G\Gamma\rho_0u_i\prez L^6+O(\epsilon^5).
\label{eq:divA} 
\end{eqnarray}
Inserting Eqs.~(\ref{eq:54}) and (\ref{eq:e36.2}) into this equation,
we find
\begin{eqnarray}
\frac{\partial_i \pret Q}{3\Gamma+2}
&=&\frac{9\Gamma^2}{4a_f^2(3\Gamma+2)\,\prez L^6}\,
\bar{D}^j\left[\prez L^6\pret F_{ij}\right],
\label{eq:57} \\
\preth C_i
&=&\frac{3\Gamma}{4a_f^3\,\prez L^6}\,
\bar{D}^j\left[\prez L^6\pret C_{ij}\right].
\label{eq:58}
\end{eqnarray}
The latter equation implies that $\preth C_j$ is not arbitrary
but expressed in terms of $\prez L$, $\prez f_{ij}$ and $\pret C_{ij}$,
while the former equation is found to be consistent with
Eq.~(\ref{eq:Qdef}). This consistency is a result of the 
Bianchi identities.

Before closing this subsection, we note the following fact.
We have considered $u_j$ in the above, but we have not 
considered the 3-velocity $v^i$ and the shift vector 
$\beta^i$ individually. As seen from Eq.~(\ref{eq:3velocity}), 
we have $v^i+\beta^i=u^0\gamma^{ij}u_j$. 
Thus, once we fix the shift vector under the assumption 
$\beta^i=O(\epsilon)$, $v^i$ is immediately determined.
The reason that we have not bothered to do this is
that it was unnecessary to fix the shift vector for the 
derivation of all the other quantities in order to 
derive their leading-order terms. 
Geometrically, $u_j$ does not depend
on the choice of the shift vector because it describes
the components projected on to the hypersurfaces of constant $t$.
The same is true for all the other variables, except for $v^i$.
This shift vector independence is probably a result 
of the localness of the gradient expansion.

\subsection{The solution to $O(\epsilon^2)$ in the gradient expansion}
Now we consider the general solution valid to $O(\epsilon^2)$
in the gradient expansion. As we have seen in the previous subsection,
among the basic variables we have introduced,
the only quantities whose leading-order terms are
lower than second order in $\epsilon$ are $\psi$ and $\tilde\gamma_{ij}$,
that is, the spatial metric. Hence, we have to
evaluate the next order terms of these variables.

To find the next order solutions for $\psi$ and $\tilde\gamma_{ij}$,
we now have to specify the shift vector. For simplicity,
we choose $\beta^i$ to be $O(\epsilon^3)$. Under this assumption,
the solutions are easily derived.
We note that
this choice includes the case of comoving coordinates,
$v^i=0$, as seen from Eq.~(\ref{eq:ordercount}).

First, we consider $\psi$. From Eq.~(\ref{eq:25}), we have
\begin{equation}
\frac{6\partial_t{\psi}}{\psi}-3\frac{\dot{a}}{a}\chi=O(\epsilon^4).
\label{eq:59}
\end{equation}
Substituting the solution~(\ref{eq:chisol}) for $\chi$,
this equation is readily integrated, and we obtain
\begin{equation}
\psi=\prez L-\frac{\prez L\,\pret Q}{6\Gamma}t^{2-\frac{4}{3\Gamma}}
+\pret L(x^i)+O(\epsilon^4),
\label{eq:60}
\end{equation}
where $\pret L$ is an arbitrary function of the spatial coordinates.
Without loss of generality, we can absorb $\pret L$ into
$\prez L$. Thus we obtain the final expression for $\psi$ as
\begin{equation}
\psi=\prez L(x^i)-\frac{\prez L(x^i)\,\pret Q(x^i)}{6\Gamma}\,
t^{2-\frac{4}{3\Gamma}}
+O(\epsilon^4).
\label{eq:psifinal}
\end{equation}

Turning to $\tilde\gamma_{ij}$, from Eq.~(\ref{eq:26}), we have 
\begin{equation}
\partial_t\tilde{\gamma}_{ij}=-2\tilde{A}_{ij}+O(\epsilon^4).
\label{eq:61}
\end{equation}
Substituting the solution~(\ref{eq:53}) for $\tilde A_{ij}$,
this equation gives
\begin{equation}
H_{ij}=-2\pret F_{ij}\int^t \frac{dt}{a^3(t)}\int^t a(t)dt
-2\pret C_{ij}\int^t\frac{dt}{a^3(t)}+O(\epsilon^4),
\label{eq:62}
\end{equation}
The integrals are easily evaluated, and we find
\begin{equation}
H_{ij}=-\frac{\pret F_{ij}}{a^2_f}\frac{9\Gamma^2}{9\Gamma^2-4}
t^{2-\frac{4}{3\Gamma}}-\frac{2\Gamma\pret C_{ij}}{a_f^2(\Gamma-2)}\,
t^{1-\frac{2}{\Gamma}}+\pret f_{ij}(x^k)+O(\epsilon^4),
\label{eq:63}
\end{equation}
where $\pret f_{ij}$ is an arbitrary function of the
 spatial coordinates. The final expression for $\tilde\gamma_{ij}$
is given by
\begin{eqnarray}
\tilde\gamma_{ij}
&=&\prez f_{ij}(x^k)
-\frac{\pret F_{ij}(x^k)}{a^2_f}\frac{9\Gamma^2}{9\Gamma^2-4}
t^{2-\frac{4}{3\Gamma}}-\frac{2\Gamma\pret C_{ij}(x^k)}{a_f^2(\Gamma-2)}\,
t^{1-\frac{2}{\Gamma}}+O(\epsilon^4)
\label{eq:tgamfinal} \\
&=&\prez f_{ik}\left(\delta^k_j-
\pret F^k{}_{j}\,\frac{9\Gamma^2}{a_f^2(9\Gamma^2-4)}\,
t^{2-\frac{4}{3\Gamma}}
-\pret C^k{}_{j}\,\frac{2\Gamma}{a_f^2(\Gamma-2)}\,
t^{1-\frac{2}{\Gamma}}\right)+O(\epsilon^4)\,,
\nonumber
\end{eqnarray}
where we have absorbed $\pret f_{ij}$ into $\prez f_{ij}$.
Here, it is important to note that $\prez f_{ij}$ is not completely
arbitrary, but its determinant must be unity:
$\det(\prez f_{ij})=1+O(\epsilon^4)$.

\subsection{Summary of the general solution}
We now have the general solutions valid through $O(\epsilon^2)$
for all the physical quantities on the uniform Hubble slicing,
under the condition $\beta^i=O(\epsilon^3)$.
We list them below:
\begin{eqnarray}
\alpha
&=&1+\chi=1-\pret Q\,\frac{3\Gamma-2}{3\Gamma}\,t^{2-\frac{4}{3\Gamma}}\,,
\label{eq:D1} \\
\psi
&=&\prez L\left(1-\pret Q\,\frac{1}{6\Gamma}t^{2-\frac{4}{3\Gamma}}\right)
=\prez L\left(1-\frac{\Gamma}{16a_f^2}R\left[L^4f\right]
\,t^{2-\frac{4}{3\Gamma}}\right)
\,,
\label{eq:D2} \\
\tilde{\gamma}_{ij}&=&
\prez f_{ik}\left(\delta^k_j-
\pret F^k{}_{j}\,\frac{9\Gamma^2}{a_f^2(9\Gamma^2-4)}\,
t^{2-\frac{4}{3\Gamma}}
-\pret C^k{}_{j}\,\frac{2\Gamma}{a_f^2(\Gamma-2)}\,
t^{1-\frac{2}{\Gamma}}\right)\,,
\label{eq:D3} \\
\tilde{A}_{ij}&=& \pret F_{ij}\,\frac{3\Gamma}{a_f^2(3\Gamma+2)}\,
t^{1-\frac{4}{3\Gamma}}
+\pret C_{ij}\,\frac{1}{a_f^3}\,t^{-\frac{2}{\Gamma}},
\label{eq:D4} \\
\cr
\delta&=&\pret Q\,t^{2-\frac{4}{3\Gamma}}\,,
\label{eq:delrho}\\
u_i&=&\prez L^{-6}\,\bar{D}^j\left[\prez L^6\pret F_{ij}\right]\,
\frac{9\Gamma^2}{4a_f^2(3\Gamma+2)}\,t^{3-\frac{4}{3\Gamma}}
+\prez L^{-6}
\bar{D}^j\left[\prez L^6\pret C_{ij}\right]\,\frac{3\Gamma}{4a_f^3}\,
t^{2-\frac{2}{\Gamma}}\,,
\label{eq:D5}  
\end{eqnarray}
where the freely specifiable functions are $\prez L$, $\prez f_{ij}$
and $\pret C_{ij}$. The functions $\pret F_{ij}$ and $\pret Q$ are given 
by Eqs.~(\ref{eq:Fijdef}) and (\ref{eq:Qdef}), respectively, as functions of
$\prez L$ and $\prez f_{ij}$. Note that 
$\prez L^4\pret F_{ij}$ is the traceless part of 
$R_{ij}\left[L^4f\right]$, i.e., the traceless part of
the Ricci tensor of the metric $\prez L^4\prez f_{ij}$, as seen from
its definition (\ref{eq:Fijdef}).

To clarify the physical role of these freely specifiable functions,
let us first count the degrees of freedom. 
Because the determinant of
$\prez f_{ij}$ is unity, it has 5 degrees of freedom,
and because $\pret C_{ij}$ is traceless, it also has 5 degrees of freedom.
Thus, together with the degree of freedom of $\prez L$,
 the total number is $1+5+5=11$, while the number of the true physical 
degrees of freedom
is $4+2+2=8$, where $4$ are for the fluid (1 from the density and 3 from
the 3-velocity) and $2+2$ are for the gravitational waves
 (2 from the metric and 2 from the extrinsic curvature). This implies
that there remain 3 gauge degrees of freedom.
It is easy to understand that these 3 gauge degrees come from
spatial covariance, that is, from the gauge freedom of
purely spatial coordinate transformations,
$x^i\to \bar x^i=f^i(x^j)$. Thus we may regard
$\prez f_{ij}$ as containing these 3 gauge degrees of freedom. 

To summarize, the degrees of freedom contained in the
freely specifiable functions can be interpreted as
\begin{eqnarray}
\prez L~ &\cdots&\ 1=1~\mbox{(density)}\,,
\nonumber\\
\prez f_{ij}~&\cdots&\ 5=3~\mbox{(gauge)}~+2~\mbox{(GWs)}\,,
\nonumber\\
\pret C_{ij}~&\cdots&\ 5=3~\mbox{(velocity)}~+2~\mbox{(GWs)}\,.
\label{eq:dof}
\end{eqnarray}
To fix the gauge completely, we therefore must impose 
3 spatial gauge conditions on $\prez f_{ij}$. 
If $\prez f_{ij}$ were not a metric, we could impose the transversality
(divergence-free) condition on it. However, this cannot be done, because
$\prez f_{ij}$ is the metric and hence, any covariant derivative of it vanishes
identically. (This is the very nature of any metric tensor.)

However, in the case of linear perturbation theory, or even in the case of
a higher-order perturbation theory, this difficulty disappears because of
the presence of the background metric. In such a case, we may impose
the transversality condition on the full (nonlinear) metric with respect 
to the background metric to single out the gravitational wave degrees of
freedom. In fact, the most commonly used conditions are those realized
by imposing transversality with respect to the flat Cartesian metric,
\begin{eqnarray}
\delta^{kj}\partial_k\prez f_{ij}\equiv\partial^j\prez f_{ij}=0\,.
\label{eq:ttgauge}
\end{eqnarray}

At this stage, it is useful to clarify 
the correspondence between the degrees of freedom
counted in Eq.~(\ref{eq:dof}) and those of the linear theory.
{}From the time dependence associated with $\prez L$, $\prez f_{ij}$
and $\pret C_{ij}$, we can identify $\prez L$ with the growing adiabatic
mode of density perturbations, the 2 degrees in $\prez f_{ij}$ with the growing modes of gravitational waves, the 3 degrees in $\pret C_{ij}$
with 1 decaying scalar and 2 vector (vorticity) modes, 
and the remaining 2 in $\pret C_{ij}$
with the decaying gravitational wave modes. 

Now, returning to the full nonlinear theory, it is generally impossible
to identify uniquely the gravitational wave degrees of freedom in
the metric, because of general covariance. Nevertheless, if we focus
on the extrinsic curvature $K_{ij}$, since its transverse-traceless
part may be defined unambiguously, we may be able to single out the
gravitational wave (tensor) degrees of freedom.

To show that this is indeed the case, let us consider
the momentum constraint~(\ref{eq:divA}),
\begin{equation}
\bar{D}^j\left[\prez L^6\tilde{A}_{ij}\right]
=8\pi G\Gamma\rho_0u_i\prez L^6+O(\epsilon^5).
\nonumber
\end{equation}
If we require the tensor modes to be transverse-traceless,
an apparent candidate is the part of $\prez L^6\tilde A_{ij}$ that
does not contribute to the right-hand side of the momentum constraint.
In other words, we identify the transverse-traceless (TT) part of
 $\prez L^6\tilde A_{ij}$
with respect to the metric $\prez f_{ij}$ as the tensor modes.
Then, the question is if the TT part can be
uniquely determined.
Fortunately, it is known that this is possible~\cite{York:1973ia}.
Here we recapitulate the result briefly. For any symmetric
traceless second rank tensor $X_{ij}$ (with respect to $\tilde\gamma_{ij}$
in our case), we have the decomposition
\begin{eqnarray}
X_{ij}&=&\tilde D_iW_j+\tilde D_iW_i
-\frac{2}{3}\tilde\gamma_{ij}\tilde D^kW_k+X^{TT}_{ij}
\nonumber\\
&\equiv&(\Lambda W)_{ij} +X^{TT}_{ij}\,,
\end{eqnarray}
where $W_i$ satisfies the second-order differential equation
\begin{eqnarray}
(O\,W)_i\equiv -\tilde{D}^j(\Lambda W)_{ij}=-\tilde{D}^j(X_{ij})\,.
\label{eq:diffope}
\end{eqnarray}
York showed that the solution can be determined uniquely~\cite{York:1973ia}.
He then further showed that this decomposition is conformally invariant.
That is, for $\hat\gamma_{ij}=\phi^4\tilde\gamma_{ij}$, we have
$\hat X_{ij}=\phi^{-2}X_{ij}$, and therefore as the TT part,
\begin{eqnarray}
\hat X^{TT}_{ij}=\phi^{-2}X^{TT}_{ij}\,.
\end{eqnarray}
Thus, identifying $X_{ij}$ with $\prez L^6\tilde A_{ij}$, we immediately
see that the transverse parts of $\prez L^6\,\pret F_{ij}$
and $\prez L^6\,\pret C_{ij}$ represent the gravitational wave modes.
Furthermore, from their time dependence, it is clear that the former 
is the growing mode while the latter is the decaying mode,
as in the case of linear theory.

Here, it may be worth noting that the TT part of 
$\prez L^6\,\pret F_{ij}$ exists even if 
the metric $\prez f_{ij}$ is trivial,
say, $\prez f_{ij}=\delta_{ij}$. The non-vanishing TT part 
in this case is due to nonlinear 
interactions of the scalar adiabatic growing mode, as can
be seen from Eq.~(\ref{eq:Fijdef}).

Finally, let us comment on the vector mode. It is contained
only in $\pret C_{ij}$, as can be seen by considering the vorticity
conservation law, which is valid for any perfect fluid~\cite{Hawking:1973uf}.
It can be shown that the time dependence
 $t^{2-\frac{2}{\Gamma}}\propto a^{3(\Gamma-1)}$ associated
with $\pret C_{ij}$ is
precisely that resulting from the vorticity conservation.

\section{Coordinate transformation to synchronous coordinates}
\label{sec:synchronous}

Comer et al.~\cite{Comer:1994np} obtained the metric to $O(\epsilon^2)$ 
in the synchronous coordinates. However, the result they obtained [Eq.~(2.14)
in their paper] has a rather different form from ours, partly because of the
difference in the choice of coordinates and partly because of the fact
that their result is not the general solution.
Here, we seek a coordinate transformation from our coordinates to
the synchronous coordinates and derive
an explicit relation between their solution and ours.

Let us denote the coordinates in the uniform Hubble slicing by
$\{x^{\mu}\}$ and those in the synchronous gauge by $\{\bar{x}^{\mu}\}$,
and the corresponding spacetime metrics by
$g_{\mu\nu}(x)$ and $\bar g_{\mu\nu}(\bar x)$, respectively.
The solution obtained by Comer et al. is
\begin{eqnarray}
\bar{g}_{ij}
&=&t^{\frac{4}{3\Gamma}}\Biggl[h_{ij}+
\Bigl(-\frac{3\Gamma^2}{4(9\Gamma-4)}t^{2-\frac{4}{3\Gamma}}
+\gamma^{\prime}_2t^{-1}+\gamma_2\Bigr)h_{ij}R[h]
\nonumber \\
& & \quad\qquad +\Bigl(-\frac{9\Gamma^2}{9\Gamma^2-4}t^{2-\frac{4}{3\Gamma}}
+\beta^{\prime}_{2}t^{1-\frac{2}{\Gamma}}+\beta_2\Bigr)
\Bigl(R_{ij}[h]-\frac{h_{ij}}{3}R[h]\Bigr)\Biggr],
\label{eq:Synsol1}
\end{eqnarray}
where 
\begin{eqnarray}
h_{ij}=\prez L^4\,\prez f_{ij}\,,
\label{hijdef}
\end{eqnarray}
and $R_{ij}[h]$ and $R[h]$ are the Ricci tensor and Ricci scalar,
respectively, of the metric $h_{ij}$, and $\beta_2$, $\beta_2'$, 
$\gamma_2$ and $\gamma_2'$ are arbitrary constants. Here and in
the rest of this section, we set $a_f=1$, in accordance with
the convention of \cite{Comer:1994np}.
We note that the terms containing $\beta_2$ and $\gamma_2$
can be absorbed into $h_{ij}$, and hence we may effectively set
these constants to zero without loss of generality. Thus
the metric we consider is
\begin{eqnarray}
\bar{g}_{ij}
&=&t^{\frac{4}{3\Gamma}}\Biggl[h_{ij}+
\left(-\frac{3\Gamma^2}{4(9\Gamma-4)}t^{2-\frac{4}{3\Gamma}}
+\gamma^{\prime}_2t^{-1}\right)h_{ij}R[h]
\nonumber\\
&&\qquad\qquad
+\left(-\frac{9\Gamma^2}{9\Gamma^2-4}t^{2-\frac{4}{3\Gamma}}
+\beta^{\prime}_{2}t^{1-\frac{2}{\Gamma}}\right)
\left(R_{ij}[h]-\frac{h_{ij}}{3}R[h]\right)\Biggr].
\label{eq:Synsol2}
\end{eqnarray}
It should be noted that the term $\gamma'_2t^{-1}$
in the first line of the above is a gauge mode
that exists in the synchronous gauge,
corresponding to the shift of the time slice $t\to t+\delta t(\bar{x}^i)$.

Now consider the coordinate transformation
\begin{eqnarray}
x^\mu\to \bar x^\mu=f^\mu(x)\,.
\label{coordtrans}
\end{eqnarray}
Under this transformation, the metric is transformed as
\begin{eqnarray}
g_{\mu\nu}(x)
=\frac{\partial \bar x^\alpha}{\partial x^\mu}
 \frac{\partial \bar x^\beta}{\partial x^\nu}\bar g_{\mu\nu}(\bar x)\,.
\end{eqnarray}
For $\bar g_{\mu\nu}$ being the metric in the synchronous coordinates,
that is, for $\bar g_{00}=-1$ and $\bar g_{0i}=0$,
we have
\begin{eqnarray}
g_{00}(x)&=&-(1+2\chi)+O(\epsilon^4)
=-\frac{\partial f^0}{\partial t}
 \frac{\partial f^0}{\partial t}
+\frac{\partial f^i}{\partial t}
 \frac{\partial f^j}{\partial t}\bar g_{ij}(\bar x)\,,
\nonumber\\
g_{k0}(x)&=&\beta_k=O(\epsilon^3)
=-\frac{\partial f^0}{\partial t}\frac{\partial f^0}{\partial x^k}
+\frac{\partial f^i}{\partial t}\frac{\partial f^j}{\partial x^k}
\bar g_{ij}(\bar x)\,.
\nonumber\\
g_{kl}(x)&=&
-\frac{\partial f^0}{\partial x^k}\frac{\partial f^0}{\partial x^l}
+\frac{\partial f^i}{\partial x^k}\frac{\partial f^j}{\partial x^l}
\bar g_{ij}(\bar x)\,.
\label{eq:transtosync}
\end{eqnarray}

Since the two metrics should coincide
in the limit $\epsilon\to0$,
and the non-trivial corrections are $O(\epsilon^2)$ in both metrics,
we may choose the coordinate transformation to be of the form
\begin{eqnarray}
f^\mu(x)=x^\mu+ F^\mu(x)\,,
\label{eq:transfunc}
\end{eqnarray}
where $F^\mu=O(\epsilon^2)$. Here we have assumed that the spatial
coordinates also coincide in the limit $\epsilon\to0$, although
there always exist 3 purely spatial coordinate
degrees of freedom, as discussed near Eq.~(\ref{eq:ttgauge}).
Inserting the above into the first two relations in Eq.~(\ref{eq:transtosync}), we readily
find
\begin{eqnarray}
\frac{\partial F^0}{\partial t}=\chi
\label{Fzero}
\end{eqnarray}
and
\begin{eqnarray}
a^2h_{ij}\frac{\partial F^j}{\partial t}=\frac{\partial F^0}{\partial x^i}
\quad
\leftrightarrow
\quad
\frac{\partial F^i}{\partial t}
=a^{-2}h^{ij}\frac{\partial F^0}{\partial x^j}\,,
\label{Fieq}
\end{eqnarray}
while the last in Eq.~(\ref{eq:transtosync}) gives
\begin{eqnarray}
g_{ij}(x)&=&\bar g_{ij}(\bar x)+O(\epsilon^3)=\bar g_{ij}(x)+
F^0(x)\partial_t\bar g_{ij}(x)+O(\epsilon^3)
\nonumber\\
&=&\bar g_{ij}(x)+
2H(t)\,F^0(x)g_{ij}(x)+O(\epsilon^3)\,.
\label{gijtrans}
\end{eqnarray}

Here it is important to note that this coordinate transformation
of the spatial metric is a conformal transformation to $O(\epsilon^2)$.
Thus it leaves the unit determinant part of the metric, $\tilde\gamma_{ij}$,
and its time derivative (or the traceless part of the extrinsic curvature),
$\tilde A_{ij}$, unchanged. Recalling that the TT part of 
$\tilde A_{ij}$ with respect to the metric $\tilde\gamma_{ij}$ is
uniquely determined and is invariant under a conformal transformation
of the metric, up to conformal rescaling of the TT part, it seems
reasonable to conjecture that the tensor modes are time-slice independent
to $O(\epsilon^2)$ for a wide class of time-slicings, though the above
result is valid only for the transformation between the synchronous and
uniform Hubble slices, in the strict sense.

Equations (\ref{Fzero}) and (\ref{Fieq}) determine
the coordinate transformation.
Integrating Eq.~(\ref{Fzero}), we find
\begin{eqnarray}
F^0&=&\int^t\chi\,dt
=-\frac{3\Gamma-2}{9\Gamma-4}\pret Q\,t^{3-\frac{4}{3\Gamma}}
+\pret E({x}^i)
\label{eq:Fzerosol}\\
&=&-\frac{3\Gamma^2(3\Gamma-2)}{8(9\Gamma-4)}\,R[h]\,t^{3-\frac{4}{3\Gamma}}
+\pret E({x}^i)\,,
\nonumber
\end{eqnarray}
where $\pret E({x}^i)$ is a function of the 
spatial coordinates (determined below),
and the second equality follows from Eq.~(\ref{eq:Qdef}). 
Substituting this into $F^0$ in Eq.~(\ref{gijtrans}) and 
noting Eq.~(\ref{eq:Synsol2}), we find the metric in
the uniform Hubble slicing to be
\begin{eqnarray}
g_{ij}(x)
&=&t^{\frac{4}{3\Gamma}}
\Biggl[h_{ij}\left(1-\frac{\Gamma}{4}R[h]\,t^{2-\frac{4}{3\Gamma}}
+\left\{\gamma^{\prime}_2R[h]+\frac{4}{3\Gamma}\pret E\right\}
t^{-1}\right)
\nonumber\\
&&\qquad\qquad
+\left(-\frac{9\Gamma^2}{9\Gamma^2-4}t^{2-\frac{4}{3\Gamma}}
+\beta^{\prime}_{2}t^{1-\frac{2}{\Gamma}}\right)
\left(R_{ij}[h]-\frac{h_{ij}}{3}R[h]\right)\Biggr].
\label{gijUH}
\end{eqnarray}
Apparently, the terms in the curly brackets 
proportional to $t^{-1}$ should vanish
for uniform Hubble slicing. This condition 
determines the function $\pret E(x^i)$. In passing, we note
that the solution obtained by Comer et al.
assumes the gauge mode to be proportional to $R[h]$,
but it can be an arbitrary function of the spatial
coordinates in general.

Now, from Eqs.~(\ref{eq:D2}) and (\ref{eq:D3}),
and noting that $g_{ij}=t^{\frac{4}{3\Gamma}}\psi^4\tilde\gamma_{ij}$
and $h_{ij}=\prez L^4\prez f_{ij}$,
our solution can be expressed as
\begin{eqnarray}
g_{ij}(x)
&=&t^{\frac{4}{3\Gamma}}h_{ik}
\left(1-\frac{\Gamma}{16}R[h]\,t^{2-\frac{4}{3\Gamma}}\right)^4
\label{eq:finalgij} \\
& &\qquad\qquad \times
\left(\delta^k_j-\frac{9\Gamma^2}{9\Gamma^2-4}
\pret F^k{}_j\,t^{2-\frac{4}{3\Gamma}}
-\frac{2\Gamma}{\Gamma-2}\pret C^k{}_j\,t^{1-\frac{2}{\Gamma}}\right),
\nonumber
\end{eqnarray}
where from Eq.~(\ref{eq:Fijdef}), we have
\begin{eqnarray}
\pret F^k{}_j=\left(R^k{}_{j}[h]-\frac{1}{3}\delta^k_jR[h]\right).
\end{eqnarray}
Comparing this with Eq.~(\ref{gijUH}), we see that these two
metrics are identical, provided that we choose
\begin{eqnarray}
-\frac{2\Gamma}{\Gamma-2}\pret C^k{}_j=\beta'_2\,\,\pret F^k{}_j\,.
\label{Cijchoice}
\end{eqnarray}
Thus, we have clarified the relation between 
our solution and the solution obtained by Comer et al.~\cite{Comer:1994np}.
As is clear from this equation, their solution corresponds
to a special choice of $\pret C_{ij}$,
similar to the choice of their gauge mode mentioned above.
Although the choice of $\pret C_{ij}$ is unimportant
in the present case of a perfect fluid because 
it contains only decaying modes, 
such a special choice is not only required but also 
renders the physical meaning of $\pret C_{ij}$ unclear.


\section{Conclusion}
\label{sec:conclusion}
In this paper, focusing on the simple case of
a perfect fluid with constant adiabatic index,
we have considered nonlinear perturbations
on superhorizon scales in the context of the (spatial)
gradient expansion. We have adopted the uniform Hubble slicing
and derived the general solution valid to second order
in spatial gradients.
This general solution is found to contain 11 arbitrary
functions of the spatial coordinates.
We have identified the physical meanings of all these
functions. In particular, the relations between these degrees of freedom
and those in the linear perturbation theory have been clarified.
They are as follows:
\begin{eqnarray}
1\ &\cdots&\ \mbox{growing scalar (curvature) perturbation}\,,
\nonumber\\
1\ &\cdots&\ \mbox{decaying scalar perturbation}\,,
\nonumber\\
2\ &\cdots&\ \mbox{decaying vector (vorticity) perturbations}\,,
\nonumber\\
2\ &\cdots&\ \mbox{growing tensor perturbations}\,,
\nonumber\\
2\ &\cdots&\ \mbox{decaying tensor perturbations}\,,
\nonumber\\
3\ &\cdots&\ \mbox{spatial gauge degrees of freedom}\,.
\end{eqnarray}
In particular, we have found that the tensor modes contained in the extrinsic
curvature (the time derivative of the metric), which can be identified
using York's conformally invariant decomposition method~\cite{York:1973ia},
are determined non-locally.
Furthermore, we have found
that the tensor modes can be generated by nonlinear
interactions of the scalar growing mode, even if there exist no 
tensor growing modes, i.e., $\prez f_{ij}=\delta_{ij}$.
Obtaining a detailed understanding
of this result and its implications is left for future studies.

Then we compared our result with that obtained by Comer et al.
in the synchronous gauge~\cite{Comer:1994np} by finding the
coordinate transformation from the uniform Hubble slicing to
the synchronous (proper time) time-slicing. In carrying out the coordinate
transformation, we found the time-slice independence of
the tensor modes to $O(\epsilon^2)$.
Then, we found
that the solution obtained by Comer et al. corresponds to a special choice of
$5=1+2+2$ decaying modes 
(i.e., 1 scalar, 2 vector and 2 tensor decaying modes)
of the 11 arbitrary functions mentioned above.

Although these decaying modes are unimportant in the
present case of a perfect fluid, 
they may not be negligible in the case of a scalar field.
In particular, as mentioned in the introduction, the decaying modes 
may play an important role, even in the case of a single-component
scalar field, if the conventional slow-roll condition is violated.
In this respect, one of the advantages of the framework presented
in this paper is that it can take account of all the physical
degrees of freedom.

Finally, we note that our result contains not only nonlinear scalar modes
but also the nonlinear coupling of scalar and tensor modes.
Thus, although the present paper is limited to the case of
a simple perfect fluid, if the framework we have developed is
extended to the case of a (multi-component) scalar field,
the resulting formalism will be a powerful tool to
investigate various non-linear, non-local effects in 
inflationary cosmology, especially the deviaton from a Gaussian
behavior of curvature perturbations in inflation. 
We will attempt to develop such an extension in the near future.

\section*{Acknowledgements}

We are grateful to N. Deruelle for very useful lectures on
gradient expansion and for many fruitful discussions.
We also thank an anonymous referee for valuable comments
which helped us to improve the paper.
This work was supported in part by 
the Monbu-Kagakusho 21st century COE Program 
``Center for Diversity and Universality in Physics",
and by JSPS Grants-in-Aid for Scientific Research 
(S) No.~14102004, (B) No.~17340075, and (A) No.~18204024.


\appendix
\section{Order counting}

Here we demonstrate explicitly the order counting
of the basic variables given in Eq.~(\ref{eq:ordercount}).
This is essentially a recapitulation of the analysis
given in Ref. \cite{Shibata:1999zs}.

In the $(3+1)$-decomposition, $u^0$ is represented as
\begin{eqnarray}
u^0&&=[\alpha^2-(\beta_k+v_k)(\beta^k+v^k)]^{-1/2}.
\nonumber 
\end{eqnarray}
{}Then, from the first two basic assumptions in~(\ref{assumptions}),
i.e., $v^i=O(\epsilon)$, $\beta^i=O(\epsilon)$,
we have
\begin{eqnarray}
u^0=\frac{1}{\alpha}+O(\epsilon^2).
\label{eq:ap1}  
\end{eqnarray}
This equation and Eq.~(\ref{eq:Edef}) yield
\begin{eqnarray}
E=(\rho+P)(\alpha u^0)^2-P=\rho+O(\epsilon^2).
\label{eq:ap2}
\end{eqnarray}
Thus, from the Hamiltonian constraint~(\ref{eq:hamconst}), we obtain
\begin{eqnarray}
H^2&=&\frac{8\pi G\rho}{3}+O(\epsilon^2),
\nonumber \\ 
&=&\frac{8\pi G\rho_0(1+\delta)}{3}+O(\epsilon^2).
\label{eq:ap3}
\end{eqnarray}
Now, since $H$ is uniform, by definition, on uniform
Hubble slicing, we have
\begin{equation}
\delta=O(\epsilon^2).
\label{eq:ap4}
\end{equation}

In general, we have
\begin{eqnarray}
\psi=O(1)\,.
\label{eq:ap16}
\end{eqnarray}
{}Also, from Eq.~(\ref{eq:5}), we have 
\begin{equation}
\partial_t w+6\frac{\partial_t{\psi}}{\psi}w
+\frac{\partial_t{\delta}}{\Gamma(1+\delta)}w
+6\frac{\partial_k \psi}{\psi}wv^k+
\frac{w\rho_0\partial_k \delta}{\Gamma \rho}v^k
+w\partial_k v^k+(\partial_k w)v^k=0\,,
\label{eq:ap5}
\end{equation}
where $w\equiv \alpha u^0$. Then Eq.~(\ref{eq:ap1}) gives
$w=1+O(\epsilon^2)$, and Eq.~(\ref{eq:ap5}) gives
\begin{equation}
\partial_t{\psi}=O(\epsilon^2)\,.
\label{eq:timepsi}
\end{equation}

{}From Eq.~(\ref{eq:Sdef}), we see
\begin{eqnarray}
S_{ij}-\frac{\gamma_{ij}}{3}S^k{}_k=O(\epsilon^2)\,.
\end{eqnarray}
Therefore, from Eq.~(\ref{eq:dotA}) we have
\begin{eqnarray}
\partial_t \tilde{A}_{ij} =\alpha K\tilde{A}_{ij}+O(\epsilon^2).
\label{eq:ap7}
\end{eqnarray}
The uniform Hubble slicing condition~(\ref{eq:alpha}), together with Eqs. (\ref{eq:ap16}) and (\ref{eq:timepsi}), gives
\begin{eqnarray}
\chi=\alpha-1=O(\epsilon^2).
\label{eq:ap8}
\end{eqnarray}
Then, Eq.~(\ref{eq:ap7}) becomes
\begin{eqnarray}
\partial_t \tilde{A}_{ij}=-3\frac{\dot{a}}{a}\,\tilde{A}_{ij}+O(\epsilon^2)\,.
\label{eq:ap9}
\end{eqnarray}
The lowest-order homogeneous solution of this equation is 
$\tilde{A}_{ij}\propto a^{-3}$, which could be 
$O(\epsilon)$. As discussed in the text, however,
we assume 
\begin{eqnarray}
\partial_t \tilde{\gamma}_{ij}=O(\epsilon^2).
\label{eq:ap10}
\end{eqnarray}
This assumption, together with Eq.~(\ref{eq:dotgamma}),
implies that the $O(\epsilon)$ part of $\tilde{A}_{ij}$ is absent.
We therefore have 
\begin{eqnarray}
\tilde{A}_{ij}=O(\epsilon^2)\,.
\end{eqnarray}

Finally, from Eq.~(\ref{eq:momconst}), we have 
\begin{eqnarray}
u_j=u^0(v_i+\beta_i)=O(\epsilon^3),
\label{eq:ap11}
\end{eqnarray}
and inserting this into the expression for $u^0$
given in Eq.~(\ref{eq:4velocity}) gives 
\begin{eqnarray}
w=\alpha u^0=1+O(\epsilon^6).
\end{eqnarray}



\begin{thebibliography}{18}

\bibitem{COBEDMR}
  G.~F.~Smoot {\it et al.},
  Astrophys.\ J.\  {\bf 396}, L1 (1992).
\\
  C.~L.~Bennett {\it et al.},
  Astrophys.\ J.\  {\bf 464}, L1 (1996)
  [arXiv:astro-ph/9601067].
\bibitem{WMAP3y}
  D.~N.~Spergel {\it et al.},
  arXiv:astro-ph/0603449.

\bibitem{WMAPng}
  E.~Komatsu {\it et al.},
  Astrophys.\ J.\ Suppl.\  {\bf 148}, 119 (2003)
  [arXiv:astro-ph/0302223].

\bibitem{Bardeen:1980kt}
  J.~M.~Bardeen,
  Phys.\ Rev.\ D {\bf 22} (1980) 1882.
\bibitem{Kodama:1985bj}
  H.~Kodama and M.~Sasaki,
  Prog.\ Theor.\ Phys.\ Suppl.\  {\bf 78}, 1 (1984).
\bibitem{Mukhanov:1990me}
 V.~F.~Mukhanov, H.~A.~Feldman and R.~H.~Brandenberger,
  Phys.\ Rept.\  {\bf 215}, 203 (1992).

\bibitem{Komatsu:2001rj}
  E.~Komatsu and D.~N.~Spergel,
  Phys.\ Rev.\ D {\bf 63}, 063002 (2001)
  [arXiv:astro-ph/0005036].

\bibitem{Babich:2004yc}
  D.~Babich and M.~Zaldarriaga,
  Phys.\ Rev.\ D {\bf 70}, 083005 (2004)
  [arXiv:astro-ph/0408455].

\bibitem{Cooray:2006km}
  A.~Cooray,
  arXiv:astro-ph/0610257.

\bibitem{Acquaviva:2002ud}
  V.~Acquaviva, N.~Bartolo, S.~Matarrese and A.~Riotto,
  Nucl.\ Phys.\ B {\bf 667}, 119 (2003)
  [arXiv:astro-ph/0209156].

\bibitem{Maldacena:2002vr}
  J.~M.~Maldacena,
  JHEP {\bf 0305}, 013 (2003)
  [arXiv:astro-ph/0210603].

\bibitem{NGmodels1}
  N.~Bartolo, S.~Matarrese and A.~Riotto,
  Phys.\ Rev.\ D {\bf 69}, 043503 (2004)
  [arXiv:hep-ph/0309033].
\\
  C.~Gordon and K.~A.~Malik,
  Phys.\ Rev.\ D {\bf 69}, 063508 (2004)
  [arXiv:astro-ph/0311102].
\\
  K.~Enqvist and S.~Nurmi,
  JCAP {\bf 0510}, 013 (2005)
  [arXiv:astro-ph/0508573].
\\
  D.~H.~Lyth,
  Nucl.\ Phys.\ Proc.\ Suppl.\  {\bf 148}, 25 (2005).
\\
  K.~A.~Malik and D.~H.~Lyth,
  JCAP {\bf 0609}, 008 (2006)
  [arXiv:astro-ph/0604387].
\\
  M.~Sasaki, J.~Valiviita and D.~Wands,
  Phys.\ Rev.\ D {\bf 74}, 103003 (2006)
  [arXiv:astro-ph/0607627].
\\
  J.~Valiviita, M.~Sasaki and D.~Wands,
  arXiv:astro-ph/0610001.

\bibitem{NGmodels2}
  M.~Zaldarriaga,
  Phys.\ Rev.\ D {\bf 69}, 043508 (2004)
  [arXiv:astro-ph/0306006].
\\
  N.~Arkani-Hamed, P.~Creminelli, S.~Mukohyama and M.~Zaldarriaga,
  JCAP {\bf 0404}, 001 (2004)
  [arXiv:hep-th/0312100].
\\
  X.~Chen, M.~x.~Huang, S.~Kachru and G.~Shiu,
  arXiv:hep-th/0605045.

\bibitem{Bartolo:2004if}
  N.~Bartolo, E.~Komatsu, S.~Matarrese and A.~Riotto,
  Phys.\ Rept.\  {\bf 402}, 103 (2004)
  [arXiv:astro-ph/0406398].

\bibitem{Matarrese:1997ay}
  S.~Matarrese, S.~Mollerach and M.~Bruni,
  Phys.\ Rev.\ D {\bf 58}, 043504 (1998)
  [arXiv:astro-ph/9707278].

\bibitem{SecondOrder}
  K.~Tomita,
  Phys.\ Rev.\ D {\bf 72}, 103506 (2005)
  [Erratum-ibid.\ D {\bf 73}, 029901 (2006)]
  [arXiv:astro-ph/0509518].

  K.~Tomita,
  Phys.\ Rev.\ D {\bf 72}, 043526 (2005)
  [arXiv:astro-ph/0505157].

  K.~Tomita,
  Phys.\ Rev.\ D {\bf 71}, 083504 (2005)
  [arXiv:astro-ph/0501663].

  K.~Nakamura,
  Phys.\ Rev.\ D {\bf 74}, 101301 (2006)
  [arXiv:gr-qc/0605107].

  K.~Nakamura,
  arXiv:gr-qc/0605108.

  K.~A.~Malik,
  JCAP {\bf 0511}, 005 (2005)
  [arXiv:astro-ph/0506532].

  H.~Noh and J.~c.~Hwang,
  Phys.\ Rev.\ D {\bf 69}, 104011 (2004).

\bibitem{Seery:2005wm}
  D.~Seery and J.~E.~Lidsey,
  JCAP {\bf 0506}, 003 (2005)
  [arXiv:astro-ph/0503692].


\bibitem{Lyth:2004gb}
  D.~H.~Lyth, K.~A.~Malik and M.~Sasaki,
  JCAP {\bf 0505}, 004 (2005)
  [arXiv:astro-ph/0411220].


\bibitem{Ellis:1989jt}
  G.~F.~R.~Ellis and M.~Bruni,
  Phys.\ Rev.\ D {\bf 40}, 1804 (1989).


\bibitem{Langlois:2005ii}
  D.~Langlois and F.~Vernizzi,
  Phys.\ Rev.\ Lett.\  {\bf 95}, 091303 (2005)
  [arXiv:astro-ph/0503416].

\bibitem{Lifshitz:1963ps}
  E.~M.~Lifshitz and I.~M.~Khalatnikov,
  Adv.\ Phys.\  {\bf 12}, 185 (1963).

\bibitem{Belinsky:1982pk}
  V.~a.~Belinsky, I.~m.~Khalatnikov and E.~m.~Lifshitz,
  Adv.\ Phys.\  {\bf 31}, 639 (1982).

\bibitem{antiNewton}
K.~Tomita, Prog.\ Theor.\ Phys.\ {\bf 48}, 1503 (1972).

\bibitem{Tomita:1975kj}
  K.~Tomita,
  Prog.\ Theor.\ Phys.\  {\bf 54}, 730 (1975).

\bibitem{Salopek:1990jq}
  D.~S.~Salopek and J.~R.~Bond,
  Phys.\ Rev.\ D {\bf 42}, 3936 (1990).

\bibitem{Comer:1994np}
  G.~L.~Comer, N.~Deruelle, D.~Langlois and J.~Parry,
  Phys.\ Rev.\ D {\bf 49}, 2759 (1994).

\bibitem{Deruelle:1994iz}
  N.~Deruelle and D.~Langlois,
  Phys.\ Rev.\ D {\bf 52}, 2007 (1995)
  [arXiv:gr-qc/9411040].

\bibitem{Muller:1989rp}
  V.~Muller, H.~J.~Schmidt and A.~A.~Starobinsky,
  Class.\ Quant.\ Grav.\  {\bf 7}, 1163 (1990).

\bibitem{Iguchi}
  O.~Iguchi, H.~Ishihara and J.~Soda,
  Phys.\ Rev.\ D {\bf 55} (1997) 3337
  [arXiv:gr-qc/9606012].
\\
  O.~Iguchi and H.~Ishihara,
  Phys.\ Rev.\ D {\bf 56}, 3216 (1997)
  [arXiv:gr-qc/9611047].

\bibitem{Khalatnikov:2003ac}
  I.~M.~Khalatnikov, A.~Y.~Kamenshchik, M.~Martellini and A.~A.~Starobinsky,
  JCAP {\bf 0303}, 001 (2003)
  [arXiv:gr-qc/0301119].


\bibitem{Salopek:1990mp}
  D.~S.~Salopek,
  Phys.\ Rev.\ D {\bf 43}, 3214 (1991).

\bibitem{HJHigherOrder}
  D.~S.~Salopek and J.~M.~Stewart,
  Class.\ Quant.\ Grav.\  {\bf 9}, 1943 (1992).

  J.~Parry, D.~S.~Salopek and J.~M.~Stewart,
  Phys.\ Rev.\ D {\bf 49}, 2872 (1994)
  [arXiv:gr-qc/9310020].

\bibitem{Soda:1995fz}
  J.~Soda, H.~Ishihara and O.~Iguchi,
  Prog.\ Theor.\ Phys.\  {\bf 94}, 781 (1995)
  [arXiv:gr-qc/9509008].

\bibitem{Nambu:1994hu}
  Y.~Nambu and A.~Taruya,
  Class.\ Quant.\ Grav.\  {\bf 13}, 705 (1996)
  [arXiv:astro-ph/9411013].

\bibitem{Sasaki:1998ug}
  M.~Sasaki and T.~Tanaka,
  Prog.\ Theor.\ Phys.\  {\bf 99}, 763 (1998)
  [arXiv:gr-qc/9801017].

\bibitem{Wands:2000dp}
  D.~Wands, K.~A.~Malik, D.~H.~Lyth and A.~R.~Liddle,
  Phys.\ Rev.\ D {\bf 62}, 043527 (2000)
  [arXiv:astro-ph/0003278].


\bibitem{Rigopoulos:2003ak}
  G.~I.~Rigopoulos and E.~P.~S.~Shellard,
  Phys.\ Rev.\ D {\bf 68}, 123518 (2003)
  [arXiv:astro-ph/0306620].


\bibitem{Leach:2001zf}
  S.~M.~Leach, M.~Sasaki, D.~Wands and A.~R.~Liddle,
  Phys.\ Rev.\ D {\bf 64}, 023512 (2001)
  [arXiv:astro-ph/0101406].


\bibitem{Shibata:1999zs}
  M.~Shibata and M.~Sasaki,
  Phys.\ Rev.\ D {\bf 60}, 084002 (1999)
  [arXiv:gr-qc/9905064].

\bibitem{York:1973ia}
  J.~W.~.~York,
  J.\ Math.\ Phys.\  {\bf 14}, 456 (1973).

\bibitem{Hawking:1973uf}
See e.g.,  S.~W.~Hawking and G.~F.~R.~Ellis,
``The Large scale structure of space-time,''
Section 4.1.
\end{thebibliography}
\end{document}